# Stress-testing university AI governance: A prospective method for locating policy breakpoints

Biranchi Poudyal

Faculty of Arts and Society, Charles Darwin University, Australia

ORCID: https://orcid.org/0000-0002-7210-5480

## Abstract

Universities are producing AI principles and use policies faster than they are building decision pathways for unfamiliar forms of AI agency. This study develops Institutional AI Governance Stress Testing (IAGST), a prospective documentary method for locating where publicly documented governance ceases to yield an accountable response. IAGST adapts established policy stress-testing and wind-tunneling logic. Its originality lies in combining controlled capability escalation, a frozen documentary corpus, a six-dimensional governance response chain, non-compensatory decision rules, and case-level breakpoint diagnosis. The method was demonstrated using 133 substantive public documents from five Western Australian universities and 15 quality-screened scenarios, resulting in 75 university-scenario encounters. Six cases were resolved, 14 were resolved through structured discretion, and 55 were indeterminate. Governed pathways fell from 16 of 25 augmentation cases to four delegation cases and none at autonomous substitution. The dominant weakness was not the complete absence of responsible roles: all 50 authority-gap cases named a role at only a generic level but lacked sufficient decision criteria or process. The findings show how universities can move beyond policy inventories and principal statements by testing whether authority, procedures, safeguards, and reviews remain connected as AI capabilities evolve. IAGST is a reproducible diagnostic for policy learning, not a ranking or measure of implementation.

## Introduction

Universities have responded to generative artificial intelligence (AI) with principles, guidelines, assessment rules, research-integrity advice, and increasingly broad institutional frameworks. Comparative studies reveal substantial variation in policy coverage, specificity, and regulatory stance (An et al., 2025; Dai et al., 2025; Jin et al., 2025). Yet policy presence is a weak proxy for managerial preparedness. A document may express a defensible principle while leaving unanswered who may decide an unfamiliar case, which criteria govern that decision, what process an affected person can use, and where review or remedy sits. Those omissions become more consequential as AI systems move from bounded assistance towards persistent delegation, representation, coordination, and cross-system action.

The management problem is therefore not simply whether a university has an AI policy. Whether the wider governance architecture can connect a novel capability to an accountable institutional response. Existing research on higher education policy has understandably focused on permitted use, disclosure, academic integrity, and assessment redesign (Luo, 2024; Weng et al., 2024; Xia et al., 2024). More recent work proposes principles-based institutional frameworks for responsible AI in research, along with associated changes to training, communication, infrastructure, and processes (Smith et al., 2026). These are important advances. They do not, however, provide a reproducible way to determine where a university's existing documents stop producing a sufficiently authorised, actionable, safeguarded, and reviewable pathway when the form of AI agency changes.

Strategic foresight offers a relevant but incomplete methodological starting point. Policy stress-testing, also known as wind-tunnelling, assesses whether policy options remain robust across scenarios or under disruptive conditions (Fernandes & Heflich, 2022; Government Office for Science, 2024). Anticipatory governance similarly seeks to embed foresight, experimentation, and learning within public decision systems rather than treating the future as a prediction problem (Kallo & Välimaa, 2025; Tõnurist & Hanson, 2020). In higher education, its importance for AI governance and leadership has recently been mapped, while the gap between theory and operationalisation remains evident (Baroudi, 2026). Conventional applications are commonly deliberative workshops: participants discuss how policies might perform in several futures and identify possible modifications. Such exercises generate learning but can be difficult to reproduce, compare across institutions, or audit back to a controlling documentary provision.

This paper develops Institutional AI Governance Stress Testing (IAGST) as a specialised adaptation for higher education. The stress imposed is controlled capability escalation within a recognisable activity: augmentation, delegation, and autonomous substitution. The object under test is not a single AI policy but the public governance corpus that jointly distributes authority, procedure, rights, safeguards, and revision responsibilities. The unit of analysis is a university-scenario encounter. The outcome is determined by conjunctive rules rather than an averaged readiness score, because a strong principle cannot compensate for the absence of an authorised decision-maker or an actionable route. The first strict failure within a capability family is recorded as a breakpoint, and all co-occurring failures form a diagnostic signature for reform.

The originality claimed here is therefore configurational and methodological, not the invention of scenario planning, policy stress testing, document analysis, or responsible AI principles. To the author's knowledge, prior studies of university AI governance have not combined five features in one auditable design: a frozen multi-document corpus; evidence-grounded scenarios arranged as controlled capability escalations; a governance response chain spanning recognition, authority, coherence, procedure, safeguards, and adaptation; deterministic non-compensatory outcome rules; and breakpoint diagnosis at the university-scenario level. This narrower claim also clarifies what IAGST does not establish. It does not measure implementation, organisational culture, staff competence, policy compliance, or the actual safety of an AI system. It tests the documented capacity to respond.

The empirical demonstration uses the five universities in the bounded Western Australian case set. The comparison is methodological rather than reputational: no institutional ranking or composite readiness score is produced. The study addresses three questions:

> RQ1. How can policy stress-testing be operationalised as a systematic and reproducible documentary method for university AI governance?
>
> RQ2. At what capability levels do publicly documented governance systems cease to produce a complete, accountable pathway?
>
> RQ3. Which recurrent governance failure signatures explain those breakpoints, and what do they imply for institutional policy and management?

## Conceptual framework

### *Public documentation as governance infrastructure*

Official documents are imperfect but consequential representations of institutional governance. They do not disclose informal coordination, resources, implementation quality, or behaviour in a live case (Bowen, 2009; Dalglish et al., 2020). They nevertheless stabilise categories, assign formal responsibilities, establish legitimate procedures, and specify the grounds on which decisions can be challenged. For students, staff, and external partners, the public corpus is often the accessible interface of the institution's authority. IAGST therefore defines documentary preparedness narrowly: the extent to which that corpus supports a traceable route from a specified governance challenge to an accountable response.

This focus shifts the analysis from document-by-document coverage to relationships among provisions. A privacy policy may govern data collection; an assessment procedure may govern authorship; an appeals policy may govern review; and an information-security standard may govern credentials. Each document may appear adequate in isolation, yet their combination fails to specify who owns a cross-system AI agent or how an affected person can contest its action. Preparedness is consequently relational. The relevant question is whether the applicable documents can be assembled, without speculative inference, into a coherent governance response.

### *The governance response chain*

Accountability is more than a principle attached to AI use. It is an institutional relationship in which an actor is obliged to explain and justify conduct to a forum that can question, judge, and impose consequences or correction (Bovens, 2007; Busuioc, 2021). Algorithmic-accountability scholarship similarly emphasises the need to specify what is being accounted for, by whom, to whom, and through which mechanisms (Wieringa, 2020). Responsible-AI frameworks articulate valuable commitments to fairness, transparency, privacy, and human oversight (Batool et al., 2025; Papagiannidis et al., 2025), but principles become operational only when authority, information, procedure, and intervention rights are connected.

IAGST represents this connection as a chain of governance responses with six analytically distinct links. Scope recognition asks whether the corpus recognises the actor, action, capability, and setting. Authority allocation asks whether a responsible role is not only named but also empowered by criteria and a route for review. Normative coherence asks whether current controlling documents give compatible directions. Procedural actionability asks whether an affected person can identify the next step in the order. Safeguard coverage asks whether the scenario's material rights and risks are substantively addressed. Adaptive capacity asks whether governance can detect change, escalate exceptions, learn from implementation, and revise. The chain is not a causal model, and its dimensions are not interchangeable indicators. It is a diagnostic representation of the documentary conditions required for an answer to be institutionally usable and reviewable.

This framing also resolves a familiar tension between central prescription and local discretion. Highly centralised rules may offer consistency but quickly become obsolete; unrestricted local discretion may accommodate disciplinary variation but obscure accountability. IAGST treats structured discretion as a governable middle position: a named authority applies stated criteria through a reachable procedure, with a review or appeal route. A generic instruction to consult a lecturer, faculty, committee, or service unit is not structured with discretion when the documents do not specify the decision basis or what follows. The standard is not the central control of every case. It is a reviewable institutional responsibility.

The non-compensatory logic follows directly from this chain conception. Composite maturity scores are useful when strengths can offset weaknesses, and the purpose is broad benchmarking. A governance pathway is different. If nobody is authorised to decide, a high score for normative coherence does not make the case actionable; if no review route exists, an otherwise clear rule may still fail an affected person. IAGST, therefore, treats dimensions as necessary diagnostic conditions at specified thresholds. The resulting classification is deliberately conservative, but it preserves the institutional meaning of a broken link and avoids presenting an attractive average as evidence of accountable operation.

***Anticipation, escalation, and capability-governance fit.***

Scenarios are plausible alternatives used to challenge assumptions under uncertainty, not forecasts of what will occur (Amer et al., 2013; Cordova-Pozo & Rouwette, 2023). Their policy value depends on transparent construction and a clear relationship to decisions (De Vito & Taffoni, 2026). IAGST limits speculative drift by grounding each scenario in evidence of present or emerging capability, fixing its actor, action, oversight, identity, data access, setting, benefit, risk, and boundary condition, and changing capability in three controlled steps. Augmentation retains substantive human control; delegation transfers a bounded task or interaction while a person sets goals and reviews outputs; autonomous substitution gives an AI system standing authority or permits action across contexts with little contemporaneous review.

The controlled sequence tests capability-governance fit: whether the same documentary system can continue to connect a recognisable activity to accountable institutional action as human control, identity, persistence, or system reach changes. A breakpoint is the first level at which a strict documentary failure appears. A failure signature records why: unrecognised scope, insufficiently structured authority, direct documentary conflict, a procedural dead end, a missing critical safeguard, or absence of adaptation mechanisms. This differs from a policy inventory, which records what documents mention; a maturity model, which usually aggregates indicators; and conventional wind-tunnelling, which relies primarily on participant deliberation. IAGST retains the prospective challenge of foresight while making the evidence trail and classification rules inspectable.

### *The assessment-centred inheritance*

University AI governance initially focused on assessment and academic integrity. Existing roles and processes can often absorb new tools when the central question is whether generated work is permitted, disclosed, or attributable (An et al., 2025; Dai et al., 2025; Jin et al., 2025). Yet this inheritance frames AI chiefly as content used by a human author. It is less well suited to an AI system that represents a student, negotiates with staff, coordinates a group, generates research evidence, maintains a cross-course profile, or operates across several institutional systems. Policies organised by traditional functional domains may therefore show ethical agreement while failing at their boundaries. This is the theoretical expectation tested in the empirical demonstration: documented governance should remain strongest where AI resembles an assessable tool and weaken where it becomes a delegate, counterparty, or cross-domain institutional participant.

## Method

### *Design and case boundary*

The study used a prospective documentary multiple-case design covering four public universities and one private university in Western Australia. The bounded set enabled a complete methodological demonstration within one state context; it was not selected to estimate national prevalence or to rank providers. A case was defined as the encounter between one university's frozen public governance corpus and one scenario. Five universities multiplied by 15 scenarios produced 75 cases. The protocol, codebook, scenario deck, and documentary corpus were frozen on 20 July 2026, before case evaluation began.

### *Documentary corpus*

Institutional websites were systematically searched for current public materials governing AI, academic integrity, assessment, research integrity and ethics, privacy and information management, student conduct and appeals, teaching and learning, delegation, procurement, risk, and institutional governance. The register contained 135 entries: 133 substantive documents and two absence or access-limitation records. Substantive totals were 38 for Curtin University, 17 for Edith Cowan University, 28 for Murdoch University, 30 for the University of Notre Dame Australia, and 20 for the University of Western Australia.

For each entry, the register retained title, official URL, document category and type, normative status, issuing authority, effective and review dates where stated, access date, and a local frozen copy. Evidence precedence ran from binding instruments, through institution-wide guidance and unit guidance, to other official webpages. A lower-authority source could clarify operation, but could not override an applicable binding instrument. Materials available only through authentication, informal communications, interviews, and observed practice were excluded. Negative-evidence findings required a documented search across the complete relevant corpus for the scenario's actor, action, capability, synonyms, responsible roles, and cross-references; inaccessible material was recorded as an access limitation rather than treated as proof of absence.

### *Scenario construction and quality screening*

Fifteen scenarios were organised into five capability families: assessment autonomy; representation and identity; collaborative agency; research agency; and persistent institutional integration. Each family contained an augmentation, delegation, and autonomous-substitution case. The sequence held the activity recognisable while varying the locus of control and the AI system's authority, identity, persistence, or system reach. Examples moved from AI-assisted drafting to delegated assessment production and semester-long autonomous completion; from private lecture assistance to delegated attendance and autonomous representation in an appeal; and from course-bounded memory to a cross-course profile and a standing cross-system agent.

A 27-source capability-evidence matrix supported the deck. Each scenario card fixed the actor, capability, action, human oversight, identity, data accessed, institutional domain, plausible benefit, governance risks, source justification, boundary condition, and diagnostic question. The cards underwent an internal quality screen on eight criteria: evidence base, plausibility horizon, higher education relevance, distinctiveness, governance focus, continuity of escalation, disciplined futurity, and documentary applicability. Each criterion was scored 0–2; a card required at least 12 of 16 with no zeros. All retained cards scored 15 or 16. This procedure establishes protocol conformity, not external or predictive validation, and the revised terminology deliberately avoids that overclaim.

### *Governance dimensions and decision rules*

The six dimensions operationalised the governance response chain and synthesised scholarship on policy-capacity, responsible innovation, AI governance, and accountability (Bovens, 2007; Papagiannidis et al., 2025; Stilgoe et al., 2013; Taeihagh, 2021; Wu et al., 2015). Each dimension was scored independently from 0 to 3 using explicit anchors (Table 1). A score of 3 required scenario-specific operational evidence; a score of 1 generally indicated only generic or analogical applicability. Zero denoted absence or failure, depending on the dimension. Scores were not summed into a readiness index.

**Table 1. IAGST governance response-chain dimensions and endpoint anchors**

| Dimension | Diagnostic question | Score 0 anchor | Score 3 anchor |
|---|---|---|---|
| **D1 Scope recognition** | Do documents recognise the actor, action, capability, and setting? | No applicable recognition | Explicit scenario-specific recognition |
| **D2 Authority allocation** | Is a decision-maker named and empowered with criteria? | No responsible authority named | Named authority, criteria, process, and review route |
| **D3 Normative coherence** | Are the controlling documents compatible? | Direct contradiction between current documents | Explicitly convergent rules |
| **D4 Procedural actionability** | Can an affected person follow an ordered process? | No actionable route | Complete sequence, requirements, and timing |
| **D5 Safeguard coverage** | Are the directly relevant rights and risks protected? | Critical safeguard absent | Specific safeguards, oversight, and remedy |
| **D6 Adaptive capacity** | Can governance detect and respond to change? | No review or adaptation mechanism | Periodic and event-triggered review with ownership |

**Source: Author's own work.**

A frozen decision algorithm classified each case. Documentary conflict required that the current provisions be directly incompatible, with no hierarchy or supersession rule. A case was indeterminate if scope was absent (D1 = 0), authority was absent or merely generic (D2 ≤ 1), no actionable procedure existed (D4 = 0), or a directly critical safeguard was absent (D5 = 0). Resolved through structured discretion required a delegating governance mode, D2 ≥ 2, D4 ≥ 2, and an established review or appeal pathway; where D2 equalled 2, separate case-relevant review evidence was required. Resolved required a prescribing mode, D1 = 3, D2 = 3, D3 ≥ 2, D4 = 3, and D5 = 3. Any remaining incomplete combination was an indeterminate near miss. This conjunctive design reflects the claim that a single missing operational element can break accountability chains.

Within each university-capability family, the first strict failure trigger established the breakpoint; a near miss did not move it earlier. All co-occurring strict failures were retained rather than compressed into a single score. The resulting failure signature separated the location of breakdown from its institutional cause and linked each cause to a distinct remediation problem.

The algorithm was fixed before case evaluation to prevent outcome categories from drifting towards intuitively preferred judgements. It was intentionally asymmetric: resolution required affirmative evidence of a complete route, whereas indeterminacy followed when a critical condition was absent or only generic. Documentary silence was not converted into prohibition, permission, or presumed staff discretion. This distinction matters because the dependent variable was the determinacy of public governance, not the researcher's view of whether a scenario ought to be allowed.

### *Manual coding, audit, and reflexivity*

The researcher manually coded all 75 cases and 450-dimension assessments using the frozen codebook, scenario cards, and original institutional documents. Every case record retained six scores and rationales, governance mode, appeal status, rule outcome, failure labels, a near-miss indicator, controlling documents, bounded quotations or negative-evidence statements, pinpoint locations, authority and currency checks, and official URLs. The 15-day primary evaluation phase commenced on 21 July 2026. Evidence reviews continued through 5 August, and final researcher validation was completed on 6 August 2026.

A second coder was not employed, and no intercoder-reliability claim is made. Dependability instead rested on prospective freezing, explicit anchors, deterministic rules, complete case records, and a post-coding rule-fidelity and evidence audit. The final audit asked whether each score was supported by applicable current evidence and whether the outcome followed from the dimension vector without interpretive substitution. Ambiguity was retained as indeterminate. This design makes the author's judgment visible and contestable, but it does not make that judgment independent. External replication and multi-coder testing remain necessary stages of method validation.

### *Analysis and ethics*

Analysis was descriptive and diagnostic. Frequencies summarised outcomes, strict failure triggers, near misses, breakpoints, and capability-family patterns. Dimensions describe the dataset, but were not treated as an interval-scale readiness index. To test whether the aggregate escalation pattern was confined to a small number of scenarios, score trajectories were also inspected across the 25 university-family sequences. No institutional ranking or significance test was produced. The study used public organisational documents and involved no participants or personal data.

## Results

### *Overall outcomes and the escalation cliff*

Across 75 cases, six (8.0%) were resolved, 14 (18.7%) were resolved through structured discretion, and 55 (73.3%) were indeterminate. No case met the strict definition of documentary conflict. The more informative pattern was the decline in the number of governed pathways with increasing agency (Table 2). At augmentation, 16 of 25 cases had either a prescribed or structured-discretion pathway. At the delegation, only four of 25 were resolved, and none were resolved through structured discretion. All 25 autonomous substitution cases were indeterminate.

**Table II. Governance outcomes by escalation level**

| Escalation level | Resolved | Structured discretion | Indeterminate | Documentary conflict | Total |
|---|---|---|---|---|---|
| **Level 1: augmentation** | 2 | 14 | 9 | 0 | 25 |
| **Level 2: delegation** | 4 | 0 | 21 | 0 | 25 |

| Level 3: autonomous substitution | 0 | 0 | 25 | 0 | 25 |
|---|---|---|---|---|---|
| **Total** | 6 | 14 | 55 | 0 | 75 |

**Source: Author's own work.**

The six-dimensional profiles moved in the same broad direction. From augmentation to autonomous substitution, mean scope recognition fell from 2.56 to 0.40, authority allocation from 2.48 to 1.00, normative coherence from 2.76 to 1.16, procedural actionability from 1.76 to 0.48, and safeguard coverage from 2.52 to 0.24. Adaptive capacity declined from 2.00 to 1.20. Across the 25 university-family sequences, scope recognition and adaptive capacity were non-increasing in all 25; the corresponding counts were 22 for authority, 23 for normative coherence, 21 for procedure, and 23 for safeguards. The escalation cliff was therefore distributed across capability families rather than produced by a single institutional or scenario outlier.

***Breakpoints varied by capability family.***

Assessment autonomy was the only family in which governance generally persisted through delegation: four of five Level 2 cases were resolved. Breakpoints occurred at autonomous substitution for three institutions and at augmentation for two, demonstrating both the comparative strength of assessment governance and the stringency of complete-pathway thresholds. Representation and identity failed earliest: all 15 cases, including private lecture assistance, disclosed proxy attendance, and autonomous representation in an appeal, were indeterminate. Applicable privacy, attendance, and appeals provisions existed, but they did not form a complete route for AI-mediated identity or participation.

The collaborative and research agency showed a shared pattern. Four of five augmentation-level group-assistance cases were governed through structured discretion, whereas every delegated and autonomous collaboration case was indeterminate. All five AI-assisted literature-synthesis cases had governed pathways, but every synthetic-evidence and autonomous-research case was indeterminate. Persistent institutional integration produced the cleanest boundary: all five course-level memory scenarios were resolved through structured discretion, while every cross-course profile and cross-system agent was indeterminate. Existing governance could allocate local responsibility, but it did not connect ownership across institutional domains.

***Failure signatures revealed thin rather than absent authority.***

Authority gaps were triggered in 50 cases, compared with 19 safeguard omissions, 15 scope failures, and 13 procedural dead ends (Table 3). 30 cases contained an authority gap without any other strict failure; 20 combined it with missing scope, procedure, or safeguards. The underlying score distribution sharpens this finding: none of the 75 cases scored zero for authority allocation. Every authority-gap case scored one. In other words, the dominant weakness was the lack of a responsible role. It was that a nominal role lacked sufficiently documented decision criteria, procedure, or review authority for the scenario. This is a management architecture problem, not simply a policy-presence problem.

**Table 3. Frequency and management meaning of strict failure triggers**

| Failure mode | Cases | Diagnostic implication |
|---|---|---|
| **Authority gap** | 50 | A role is nameable but lacks the required criteria and/or process; all 50 cases scored D2 = 1 |
| **Safeguard omission** | 19 | A scenario-critical protection, oversight, or remedy was absent |
| **Scope failure** | 15 | The governance corpus did not recognise the relevant action or setting |
| **Procedural dead end** | 13 | No actionable route existed for seeking or reviewing a decision |
| **Documentary conflict** | 0 | No direct contradiction between the current controlling documents |
| **Adaptation gap** | 0 | No adaptation mechanism; zero strict triggers do not imply mature adaptive capacity |

**Source: Author's own work.**

Five cases were classified as indeterminate near misses because no strict failure trigger applied, but the complete evidence threshold was not met. These cases are useful for reform because they require targeted completion rather than the creation of a new governance domain. Procedural actionability had the lowest overall mean (1.20), followed by safeguard coverage (1.39), scope recognition (1.43), authority allocation (1.64), adaptive capacity (1.72), and normative coherence (1.77). No case scored 3 for adaptive capacity. The zero count for strict adaptation gaps, therefore, does not indicate mature adaptability; it indicates that some review mechanism was usually nameable, while event triggers, ownership, and feedback into revision remained incomplete.

The combination of strict failures and near misses prevents the 73.3% indeterminate result from being read as a single undifferentiated deficit. Some cases lacked recognition of the activity; others fell within an established policy domain but stopped at nominal authority; still others almost met a prescribed or delegated pathway and needed a single missing operational element. The diagnostic value lies in these configurations. A university facing many near misses has a different reform problem from one whose scenarios repeatedly trigger combined scope, procedure, and safeguard failures, even if both have the same headline proportion of indeterminate cases.

## Discussion

### *Theoretical contribution: from policy presence to capability-governance fit*

The study's first contribution is a more demanding account of institutional preparedness. Comparative policy analyses appropriately ask whether universities mention AI, which uses they allow, and which principles recur (An et al., 2025; Dai et al., 2025; Jin et al., 2025). IAGST asks a different question: whether a documentary governance response chain remains intact as capability changes. The empirical demonstration shows why these matters are important. Normative coherence was comparatively strong, and direct conflict was absent, yet almost three-quarters of cases were indeterminate. Ethical agreement did not reliably translate into ownership, procedure, safeguards, and review.

Capability-governance fit also clarifies why an institution can appear well governed at one level and fail at the next without contradicting itself. Assessment policies often classify generated work and empower existing academic roles, thereby governing delegated production. An AI delegate attending a class, representing a student, evaluating group members, or combining data across courses changes the identity and authority relationships that those documents assume. The governance problem is not merely AI. It is a shift in who acts, whose credentials or interests are represented, which organisational boundary is crossed, and who can be called to account.

### *Methodological contribution: an auditable adaptation of policy stress-testing*

The study's second contribution is methodological. Policy stress-testing already exists as a foresight practice to probe the robustness of policies to scenarios (Fernandes & Heflich, 2022; Government Office for Science, 2024). IAGST makes four adaptations for comparative documentary research in higher education. First, it freezes the object under test and treats the entire applicable governance corpus, rather than a favoured AI policy, as the institutional response system. Second, it uses controlled capability escalation within families instead of unrelated future narratives. Third, it replaces deliberative judgements about general robustness with explicit dimensions and conjunctive decision rules. Fourth, it produces a breakpoint and failure signature that can be audited to case-level evidence and translated into specific policy work.

These adaptations create a different kind of claim from a workshop or maturity score. IAGST does not say that one university is generally ready and another is not. Nor does it claim to predict the adoption of autonomous agents. It identifies the first plausible capability condition under which current public documents fail to yield a complete response and makes the reason for this failure inspectable. Because the corpus, scenarios, anchors, and algorithms are separable, future replications can replace any component while preserving the core logic. That modularity is central to the method's usefulness and falsifiability.

### *Implications for university policy and management*

The strongest managerial implication is that universities do not primarily need another list of AI principles. They need a decision infrastructure that makes those principles executable. The 50 authority-gap cases all contained a generic role, so simply naming an office, committee, dean, lecturer, or policy owner will not resolve the dominant weakness. A governable allocation couples four elements: an authorised role, decision criteria, a reachable procedure, and a route for review. Where these elements already exist separately, reform may involve connective clauses, cross-references, standard request forms, and escalation rules rather than a wholesale new policy.

Failure signatures support differentiated remediation. Recurrent authority-structure gaps may justify a cross-functional AI governance forum with delegated authority and published decision criteria. Procedural dead ends call for an intake, triage, decision, documentation, and review pathway. Safeguard omissions should be linked to procurement, course design, research governance, privacy impact assessment, accessibility, and appeals rather than left in an abstract ethics statement. Scope failures require definitions that address representation, persistence, autonomous action, and cross-system integration without hard-coding a specific product. Adaptation weaknesses require scheduled review plus event triggers, assigned owners, incident learning, and a documented route from monitoring to revision.

Structured discretion deserves particular attention. Fourteen augmentation cases were governable without a fully prescribed substantive answer because authority, criteria, process, and review were sufficiently connected. This suggests that agility does not require regulatory vagueness. Universities can preserve disciplinary and local judgment while maintaining consistency and due process. Conversely, an unstructured referral may create the appearance of flexibility while transferring uncertainty and risk to students or frontline staff.

For management use, stress-testing should operate as a governance maintenance cycle rather than a one-off audit. A university can freeze a corpus, select a small scenario deck tied to its strategic technology horizon, record breakpoints, assign remediation owners, and repeat the test after policy or system changes. Near misses can be prioritised for rapid repair; recurrent multi-failure signatures may require redesign across committees or portfolios. Repetition also creates a modest but useful learning record: whether reforms move the breakpoint, whether new safeguards introduce procedural dead ends, and whether ownership remains clear when capabilities or vendors change.

The results also show an organizational inheritance. Governance performed best when AI could be interpreted through assessment and research-integrity roles, and worse when it crossed boundaries involving identity, participation, memory, data, or administration. Future AI governance should therefore be designed around capability and accountability relationships as well as traditional functional silos. This does not require centralising every decision. It requires making cross-domain ownership explicit before systems are procured or practices become routine.

***Implications for research***

IAGST opens a research programme rather than concluding one. Replications should preserve corpus freezing, evidence precedence, negative-evidence searches, scenario boundaries, scoring anchors, and rule-based outcomes. Multi-coder studies can test interpretive consistency; sensitivity analyses can examine the effects of alternative completeness thresholds; and longitudinal repetitions can determine whether documented breakpoints move after reform. External validity should be assessed by comparing documentary results with decision simulations, interviews, incident reviews, implementation audits, or observed case handling. The crucial question is whether documented capability-governance fit predicts how institutions actually respond.

Comparative research should also examine which organisational arrangements produce robust structured discretion. Candidate explanations include policy hierarchy, the location and delegated authority of AI governance bodies, the integration of privacy and academic decision processes, and the availability of accessible appeals. Such work would move beyond counting provisions towards explaining why some institutions connect domains more effectively. It may also reveal trade-offs: a highly explicit pathway can improve determinacy while narrowing legitimate local judgement, whereas broad discretion can support experimentation while weakening consistency. IAGST makes those trade-offs empirically visible without resolving them in advance.

## Limitations

Seven limitations bound the claims. First, IAGST measures public documentary preparedness at one point in time, not implementation, resources, informal coordination, culture, or behaviour in a live case. Second, public evidence may underestimate institutions whose operative procedures are internal; an access barrier is not proof of organisational absence. Third, the Western Australian case set demonstrates the method and does not represent Australian universities generally. Fourth, one researcher constructed, coded, and audited the cases, so interpretive independence and intercoder reliability are not established. Fifth, scenario quality depends on the capability evidence and boundaries fixed at the freeze date. Sixth, the dimensions and conservative conjunctive thresholds require external validation and sensitivity testing; workable improvisation may exist where the documents remain indeterminate. Seventh, the five capability families do not exhaust plausible AI futures, and autonomous-substitution scenarios are stress conditions rather than adoption forecasts.

## Conclusion

The editorial question raised by this study is simple but often avoided: when a university says it governs AI, what happens when technology no longer resembles the use case around which its documents were written? Institutional AI Governance Stress Testing provides one reproducible way to answer. It adapts established policy stress-testing to a frozen documentary corpus, controlled capability escalations, a governance response chain, non-compensatory rules, and breakpoint diagnosis. In the Western Australian demonstration, governed pathways were common during augmentation, scarce during delegation, and absent during autonomous substitution. The dominant weakness was not policy conflict or complete ownerlessness, but thin authority: roles could be named without the criteria and procedures needed to act accountably. That distinction turns a broad claim of “unpreparedness” into a practical reform agenda. Used iteratively, IAGST can help universities connect foresight to governance redesign before novel AI practices harden into unmanaged institutional dependencies.